\documentclass{aa}

\usepackage{txfonts}
\usepackage{orcidlink}
\usepackage{hyperref}

\makeatletter
\renewcommand*\aa@journalname{arXiv-only}
\renewcommand*\aa@manuscriptname{}
\renewcommand*\aa@publishlink{}
\renewcommand*\aa@numarticle{}
\renewcommand*\aa@textidlineempty{} 
\AtBeginDocument{\renewcommand*\aa@pageof{\thepage}}
\makeatother

\newcommand{\crires}{CRIRES$^+$}
\newcommand{\um}{$\mu$m}
\newcommand{\kms}{km\,s$^{-1}$}
\newcommand{\viper}{\texttt{viper}}
\newcommand{\vipere}{\texttt{vip\`ere}}
\newcommand{\esorex}{\texttt{esorex}}
\newcommand{\crtwores}{\texttt{cr2res}}

\begin{document}

\title{An archive of reduced and telluric-corrected CRIRES$^+$ L- and M-band spectra with slit-tilt and wavelength calibrations}
\titlerunning{CRIRES$^+$ L/M-band spectral archive}

\author{Thomas Marquart\inst{1}\,\orcidlink{0000-0002-2244-9920}
\and	Alexis Lavail\inst{2}\,\orcidlink{0000-0001-8477-5265}}
\authorrunning{T.\ Marquart \& A. Lavail}

\institute{Department of Physics and Astronomy, Uppsala University,
  Box 516, SE-75120 Uppsala, Sweden \\
  \email{thomas.marquart@astro.uu.se}
 \and Univ. de Toulouse, CNRS, IRAP, 14 avenue Belin, 31400 Toulouse, France}

\date{March 2026}

\abstract{
The high-resolution near-infrared spectrograph \crires{} at ESO VLT covers the
Y, J, H, K, L and M bands. The U-Ne and Fabry-Pérot calibration light sources, however, only work up to the K-band,
leaving the bands L and M without wavelength calibration, and without a way to
measure the inclination of the long slit relative to the detector frame.
This has made L/M data awkward to use, reflected both in the low publication fraction
 and in their absence from ESO's reduced science products.
 
To remedy this, we present here a uniformly reprocessed archive of all public \crires{} L-band
(2.8--4.2\,\um) and M-band (4.4--5.5\,\um) science observations obtained
since the instrument upgrade (September 2021) until March 2025, totalling 11\,131 raw frames
across 16 wavelength settings. We use the telluric modelling tool \viper\ 
that fits a model to the plethora of atmospheric absorption features that exist around these
wavelengths. 
To calibrate the slit tilt, we select observations of telluric standards
and measure the difference in the \viper\ wavelength solutions between the nodding A and B frames
that have the target in the lower and upper half of the slit, respectively.

We then update the static inputs to the ESO data reduction system (DRS) with the slit tilt
information and reduce the flat-fields and science data with the standard DRS recipes.
Subsequently, we derive new wavelength scales for each observation from telluric fits on the spectra 
themselves, additionally interpolating the solutions for spectra that have no tellurics from the ones that have. 
The resulting 5649 extracted, calibrated and telluric-fitted AB nod-pair spectra, spanning
156 unique targets from 68 ESO programmes, are served through an
interactive web archive at \url{https://www.astro.uu.se/crires-lm} that offers data downloads and figures for 
all datasets that allow an initial judgement of the data quality.

In addition, we provide all scripts and code, and the DRS input FITS files with the slit tilt and much improved
default wavelength solutions, for use in custom and future data reductions at the repository \url{https://github.com/ivh/CRIRES-LM}.
}


\maketitle
\nolinenumbers

\section{Introduction}
\label{sec:intro}

\crires{} is the upgraded cryogenic high-resolution infrared echelle
spectrograph at ESO's Very Large Telescope \citep{Dorn2023}. It provides
spectra from 0.95 to 5.3\,\um{} with a
resolving power of $R \approx 100\,000$ or $R \approx 50\,000$ (using respectively 
a 0.2 arcsecond or 0.4 arcsecond wide slit), recording cross-dispersed long-slit
spectra on three HAWAII2RG detectors of $2048 \times 2048$ pixels each 
with 5--9 echelle orders per chip.

While the $YJHK$ bands of \crires{} are fully reduced and calibrated by
the standard \crtwores{} pipeline \citep{cr2res} and included in ESO's
Phase~3 data releases, the thermal infrared L-band (2.8--4.2\,\um) and
M-band (4.4--5.5\,\um)\footnote{The \crires\ M-band settings have wavelength coverage
 well into the L-band, as can be seen in Fig.~\ref{fig:slittilt}.}
 settings have only the flat-field calibration lamp
available.

Therefore, two instrument properties cannot be calibrated in
L/M reductions like in the shorter bands: (1) the slit tilt, i.e.~the incline with respect to
the detector columns, which is 
wavelength-dependent and significant (up to several pixels between top and bottom
of the slit); and (2) the wavelength calibration.
As a result, L/M data are excluded from the Phase~3
archive. 

This paper describes a bulk reprocessing of all public \crires{} L/M
science data that addresses both issues and additionally provides telluric
correction. The reduced spectra are served through an interactive web
archive at \url{https://www.astro.uu.se/crires-lm/}. The reduction
code is publicly available at
\url{https://github.com/ivh/CRIRES-LM}.

In the following, we describe the methods used, assess the quality of the results
and the limitations, and highlight some examples.

\section{Data,  Methods \& Results}
\label{sec:data-methods}

We query the ESO archive for all \crires{} science frames with
wavelength settings in the L- and M-bands, covering the period from
September 2021 (first \crires{} science operations) through March 2025. The proprietary
period for \crires{} observations is 1 year, so we can only access and reduce
public data older than that.
This yields 11\,131 raw frames from 68 programmes observing 156 distinct
targets across 16 wavelength settings (Table~\ref{tab:settings}).

The observations use the standard \texttt{CRIRES\_spec\_obs\_AutoNodOnSlit}
template with ABBA nodding along the slit where the target is moved between two
distinct positions (A,B) on the slit. The nodding procedure greatly facilitates
the data reduction process throughout the near-infrared. Frames are paired into AB
nod pairs by greedy matching within each observing template sequence,
producing 5649 pairs across 412 combined observing sequences. An
additional 233 flat-field calibration epochs (deep flats with
NDIT\,$\geq$\,10) are retrieved for the same period.

\begin{table}
\caption{Wavelength settings and data volume.}
\label{tab:settings}
\centering
\begin{tabular}{lrrr}
\hline\hline
Setting & Range (nm) & Frames & Pairs \\
\hline
L3244 & 2869--3982 & 19 & 16 \\
L3262 & 2886--4003 & 972 & 498 \\
L3302 & 2923--4050 & 921 & 446 \\
L3340 & 2960--4082 & 765 & 399 \\
L3377 & 2843--4136 & 726 & 395 \\
L3412 & 2873--4177 & 61 & 36 \\
L3426 & 2886--4193 & 222 & 101 \\
\hline
M4187 & 3584--5504 & 76 & 40 \\
M4211 & 3381--5535 & 1118 & 578 \\
M4266 & 3427--5604 & 137 & 68 \\
M4318 & 3470--5154 & 1241 & 638 \\
M4368 & 3513--5211 & 4761 & 2398 \\
M4416 & 3552--5264 & 18 & 12 \\
M4461 & 3614--5314 & 74 & 39 \\
M4504 & 3870--5364 & 12 & 8 \\
M4519 & 3435--5370 & 8 & 4 \\
\hline
Total & & 11\,131 & 5649 \\
\hline
\end{tabular}
\end{table}

The reduction proceeds in four stages: slit tilt calibration from
telluric standard stars (\S\,\ref{sec:tilt}), default wavelength scale
update (\S\,\ref{sec:defaultwave}), per-observation trace adjustment and
extraction (\S\,\ref{sec:extraction}), and telluric correction
(\S\,\ref{sec:tellcorr}) with wavelength calibration.

\subsection{Slit tilt calibration}
\label{sec:tilt}

The \crires{} slit image is not aligned with the detector columns;
the resulting tilt is wavelength-dependent and, if uncorrected, causes
spectral features to shift in wavelength as a function of position along
the slit. For nodding observations, where the spectrum is extracted at
two different slit positions, ignoring the tilt introduces a systematic wavelength
offset between the A and B nod spectra. In addition, the spectral resolution
gets degraded.
In the L and M bands, the slit
tilt ranges from about $+0.01$ to $-0.09$ pixels per spatial pixel, $\approx 180$ of
which make up the full length of the slit (10\arcsec on sky). This tilt corresponds
to wavelength offsets of several \kms{} over the nod throw (typically $\approx 5\arcsec$).

\begin{figure}
  \centering
  \includegraphics[width=\columnwidth]{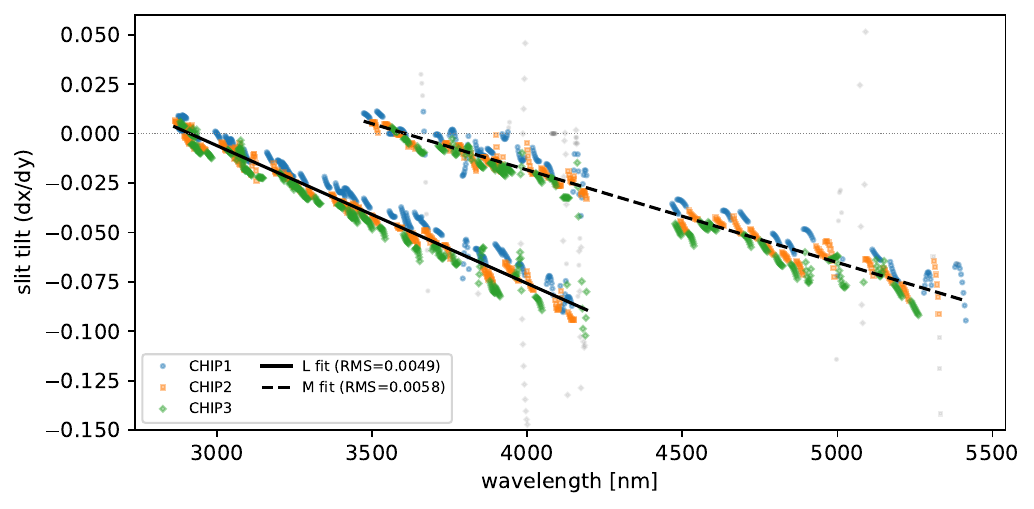}
  \caption{Slit tilt measured from the A--B wavelength difference of
    telluric standard stars, as a function of wavelength.  Different
    symbols denote the three detector chips.  Solid and dashed lines show
    the linear fits for L and M bands (Eqs.~\ref{eq:tiltL}
    and~\ref{eq:tiltM}), which interpolate across the CO$_2$ gap near
    4.2--4.5\,\um.  Grey points are 3$\sigma$ outliers excluded from the
    fit.}
  \label{fig:slittilt}
\end{figure}
\begin{table}
\caption{Telluric standard stars used for slit tilt and wavelength
  calibration.}
\label{tab:tellstds}
\centering
\begin{tabular}{llll}
\hline\hline
Settings & Star & Sp.\ type & Slit \\
\hline
L3244--L3302, L3412, L3426 & $\alpha$\,Eri & B6\,V & 0\farcs2 \\
L3340 & HD\,153426 & O9\,II & 0\farcs4 \\
L3377 & $\theta$\,Aql & B9.5\,III & 0\farcs4 \\
M4187, M4266 & HR\,7950 & B & 0\farcs2 \\
M4211 & $o$\,Hya & B9\,III & 0\farcs2 \\
M4318 & $\theta$\,Aql & B9.5\,III & 0\farcs4 \\
M4368 & $\beta$\,Ori & B8\,Ia & 0\farcs2 \\
M4416--M4504 & $\alpha$\,Eri & B6\,V & 0\farcs2 \\
M4519 & V*\,S\,CrA & T\,Tauri & 0\farcs2 \\
\hline
\end{tabular}
\tablefoot{All standards are hot stars with featureless L/M-band continua,
  except M4519 where V*\,S\,CrA is the only available target; its
  \viper{} fits are adequate but noisier.}
\end{table}

We calibrate the slit tilt using a dedicated telluric standard star for
each of the 16 wavelength settings (Table~\ref{tab:tellstds}). These are
hot B- and O-type stars whose intrinsic spectra are featureless in the
L/M bands, so the observed spectral features are purely telluric. Each
standard is observed with an AB nod pair along the slit.

We reduce each nod pair with \esorex{} \texttt{cr2res\_obs\_nodding} and
then fit the extracted A and B spectra independently with \vipere{}\footnote{\url{https://git.astro.lavail.net/alexis/vipere}},
a fork of the \texttt{viper}\footnote{\url{https://github.com/mzechmeister/viper}}
tool \citep{viper} that simultaneously fits the
telluric absorption, wavelength solution, continuum, and instrumental
profile. \footnote{In the following, to avoid confusion, we write \viper\ instead of \vipere\ since
the telluric fitting routine originates in the former.}
The key output for our purpose is the wavelength polynomial per detector order and
per nod position.

The slit tilt at each pixel is then computed as the wavelength difference
between the A and B fits, converted to a pixel shift and divided by the
measured nod throw (in pixels). The nod throw is determined by fitting
double Gaussians to the spatial profile of the combined A frame.

Collecting measurements across all settings and chips, we fit a linear
relation between tilt and wavelength for each band:
\begin{align}
  t_L(\lambda) &= -6.97 \times 10^{-5}\,\lambda + 0.203 \quad
    (\mathrm{rms} = 0.005), \label{eq:tiltL} \\
  t_M(\lambda) &= -4.68 \times 10^{-5}\,\lambda + 0.169 \quad
    (\mathrm{rms} = 0.006), \label{eq:tiltM}
\end{align}
where $\lambda$ is in nm and $t$ is the tilt in pixels per spatial pixel.
The fit uses iterative 3$\sigma$ clipping and interpolates smoothly
across the CO$_2$ fundamental absorption band near 4.2--4.5\,\um{} where
tilt measurements are unreliable. The resulting tilt values are written
into the \texttt{SlitPolyB} column of the tracing tables used for all
subsequent reductions (Fig.~\ref{fig:slittilt}).

We choose to ignore the change in tilt within each detector-order
and adopt a single value for each. While the intra-order change is clearly
visible in Fig.~\ref{fig:slittilt}, it proves difficult to derive reliably for
the orders with weak telluric features. The ``fixed tilt'' for each
order already captures the bulk of the improvement over the previous 
``no tilt'' assumption.

\subsection{Better default wavelengths}
\label{sec:defaultwave}

The \viper{} fits to the telluric standards produce
wavelength solutions. For traces where the fit converged well
(percentage RMS residual $< 10$ and wavelength shift from the pipeline
solution $< 10$\,\kms), we let the \viper{} polynomial replace the old static
 calibration in the tracing table. This applies to 225 out of
311 traces (72\%). For the remaining traces, where telluric features are weak
or block all light like in the CO$_2$ gap, we retain
the pipeline wavelength solution for now.

At this stage, we also remove some redundant duplicate traces, and partial ones
that were cut at the detector edges from the trace table files (\texttt{*\_tw.fits}).

\subsection{Calibrations: flats are all you need}
\label{sec:calibs}

\begin{figure}
  \centering
  \includegraphics[width=\columnwidth]{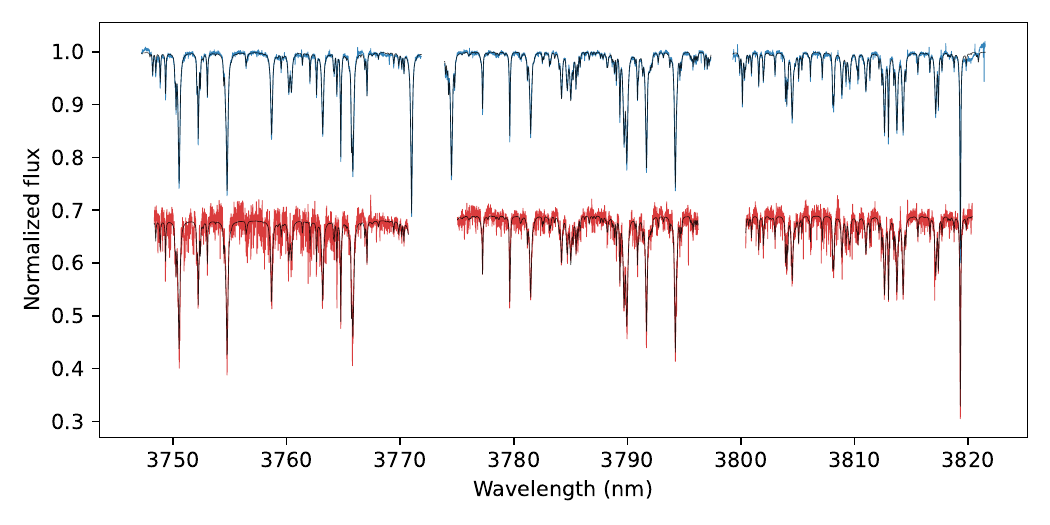}
  \caption{Extracted spectrum of $\beta$\,Pic in the M4368 setting
    (order~7, all three detector chips), reduced with flat-fielding (blue)
    and without (red, offset vertically).  The black lines show the corresponding
    \viper{} telluric model for each case. }
  \label{fig:flatcomp}
\end{figure}

Since there are no wavelength calibration lamp sources, and the detector
background is subtracted together with the sky when we difference
 the  A and B nod frames, only the flat-fields remain to be reduced and applied.
These are, however, essential for several reasons. Data with good
signal-to-noise ratio (SNR) needs flat-fields of matching depth to achieve the expected noise levels, as 
Fig.~\ref{fig:flatcomp} clearly illustrates by showing a spectrum before and after
flat-fielding. 

In addition, we pass the blaze function that \texttt{cr2res\_cal\_flat} produces
on to the science recipe. This does not fully normalize the continuum but
generally straightens it enough to make it easier for \viper\ to do its fit. Lastly,
outliers in the flat-field, with normalized sensitivity below 0.8 or above 1.2, get flagged
as bad, thus providing the bad pixel map (BPM) for the science reduction.

The \crires\ team takes deep flat-fields with high NDIT (20, 50 or 100)
from time to time. To avoid ever limiting the science products' SNR, we
download only these sets of flat-fields (not the daily ones) for all the L/M settings.
We then reduce them with \texttt{cr2res\_cal\_flat}, passing the updated
slit tilt information as input. We associate the nearest-in-time set
  of flats with each observation, and prepare the set-of-files (SOF) for the
science reduction accordingly. Each set of flat-fields has its own page 
at the archive website, and each reduced observation links to the
page of the flat-field used.

Even though the HAWAII2RG detectors are known to start deviating from linear response
already below 10\,000 ADU ($\approx $ 20\,000 electrons), we do not
apply the instrument's standard non-linearity calibration, for two reasons. First, since
\crires\ cannot illuminate its detectors fully and evenly, this calibration is taken
with a large number of flat-field frames, and still does not reach high SNR everywhere, leaving
a risk to degrade data quality. 

Second, except for very bright targets, L/M exposure times are capped by the high sky background, that 
gets removed by the subtraction of the nodding A and B frames from each other.
Thus, the exposure level of continuum and the trough of spectral lines usually
correspond only to a small fraction of the full dynamic range of the detectors.
Nevertheless, we encourage users that intend to, for example, analyse the line shapes
of the strongest spectral features in the brightest targets, to look into
applying an average non-linearity correction themselves.

\subsection{Science extraction}
\label{sec:extraction}

\crires{} has a cross-disperser wheel that selects the grating for each band.
This mechanism is not very repeatable and can leave the spectral orders shifted
vertically on
the detector by typically 5--10 pixels between the daytime calibration
and the nighttime science observation, occasionally reaching up to
${\sim}50$ pixels. The instrument therefore has a metrology routine that 
measures the positions of known lines and adjusts the wheel to the 
reference position.

Some observers, however, prefer to save on overhead time and skip the metrology,
leaving the traces offset from calibrations.
We correct for this by searching for the integer Y-shift that
maximizes the total flux contained within the order boundaries of the
reference tracing table, evaluated on the spatial profile of each raw
science frame (median along columns 500--1500). A parabolic fit around
the maximum provides sub-pixel refinement, and the median shift across
the three chips is applied to the trace polynomials before extraction.

On rare occasions it happens that the vertical offset between the flat-field
and the observation is so large, that there is no valid flat-field for the pixels
that make up one of the A or B nod positions. This data is essentially lost
and not recoverable. However, we also provide reduction runs \emph{without}
flat-fielding, which are generally inferior, but potentially useful in these cases.

This adjusted tracing table, together with the nearest-in-time master flat
and blaze function, and the science frames themselves, make up the input to
\esorex{} \texttt{cr2res\_obs\_nodding}. This recipe applies the flat-field,
subtracts the A and B frames from each other and optimally extracts the spectra, using the adjusted
traces and the slit tilt from above.
It then applies the blaze function and assigns the default wavelengths to
each spectral bin.

The number of AB pairs per observation varies greatly, and for some sequences the
time resolution is relevant, for example exoplanet transits. We therefore reduce
all data twice, once as combined average frames from all raw frames taken
by the same template, and once as single AB pairs. Both are available from
the archive website.


\begin{figure*}
  \centering
  \includegraphics[width=\textwidth]{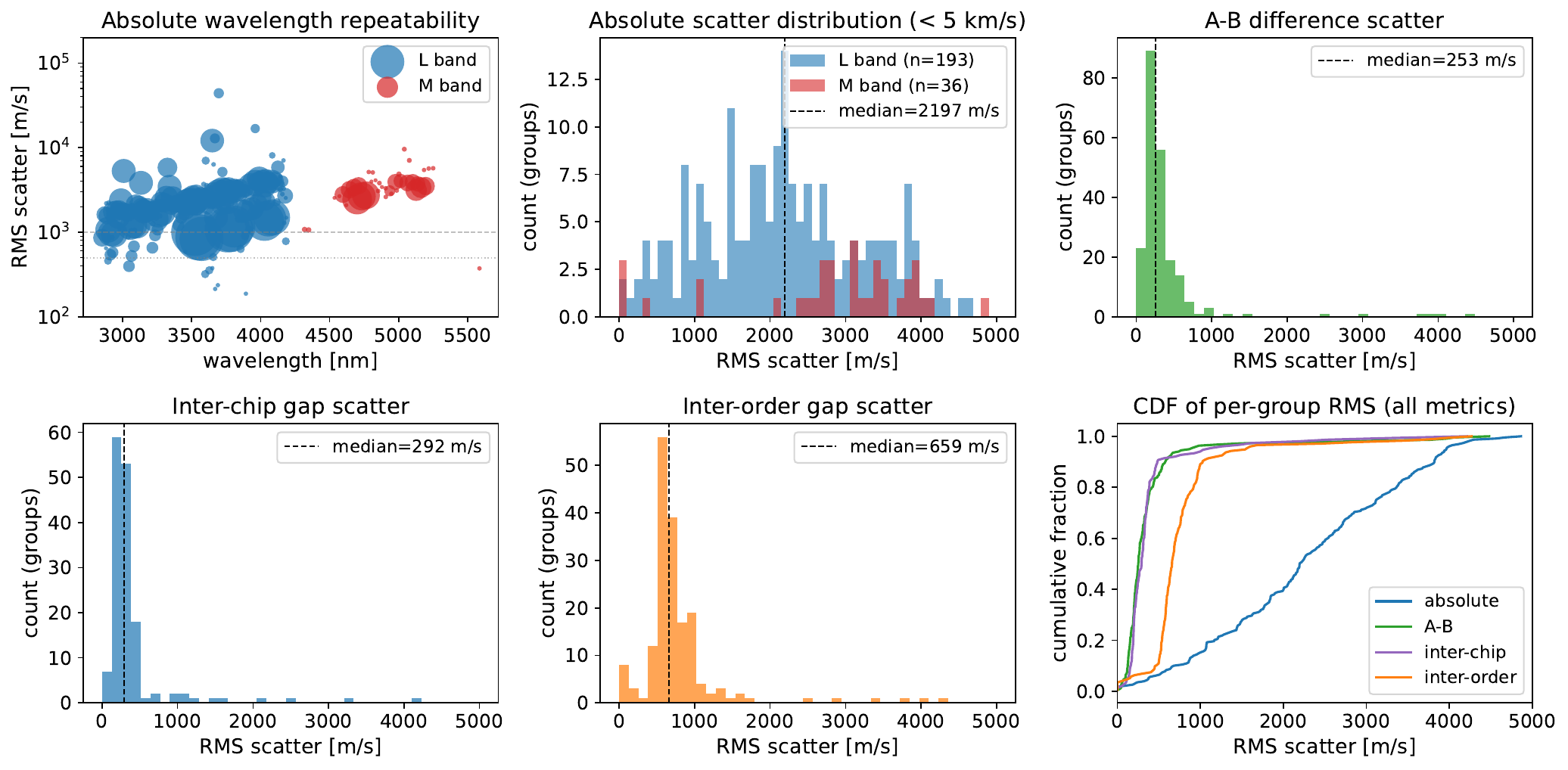}
  \caption{Wavelength precision of the \viper{} fits to science spectra,
    assessed via four metrics. \emph{Top left:} absolute repeatability of
    the wavelength at pixel~1024 per (setting, chip, order) group, with
    symbol size proportional to the number of measurements.
    \emph{Top centre:} distribution of per-group RMS for groups with
    RMS\,$<$\,5\,\kms. \emph{Top right:} A--B difference scatter, the
    tightest metric (median 249\,m\,s$^{-1}$).
    \emph{Bottom left and centre:} inter-chip and inter-order gap scatter.
    \emph{Bottom right:} cumulative distribution of per-group RMS for all
    four metrics, showing the hierarchy from A--B (tightest) to absolute
    (broadest).}
  \label{fig:wlprecision}
\end{figure*}

\subsection{Telluric correction}
\label{sec:tellcorr}

\viper{} requires pre-computed atmospheric transmission templates for each
molecule. The existing templates bundled with \viper{} covered the visible
through K~band. We therefore generate L- and M-band templates from the
\texttt{molecule\_atlas\_UVB-MIR.fits} included in the \texttt{molecfit}
v4.4.2 calibration data \citep{molecfit}, which contains per-molecule
transmission spectra from 0.29 to 30\,$\mu$m 
and the \texttt{aer\_v\_3.8} line database for a standard
Paranal atmosphere at zenith.
We extract the relevant wavelength ranges
(2.5--4.2\,$\mu$m for L, 4.2--5.5\,$\mu$m for M) and the dominant
absorbing species (H$_2$O, CH$_4$, CO$_2$, N$_2$O, CO, O$_3$) into
separate FITS tables matching the format of the existing \viper{} templates.

Each extracted science spectrum is then fitted with \viper{}, which models
the observed spectrum as a continuum polynomial multiplied by the
convolved atmospheric transmission.  The telluric transmission model is
reconstructed from the fit residuals as
$T = (\mathrm{observed} - \mathrm{residual}) / C$, where $C$ is the
fitted continuum. The corrected spectrum is then
$S_\mathrm{corr} = S_\mathrm{obs} / T$.

For orders where the \viper{} fit fails (e.g.\ the CO$_2$-saturated
region near 4.26\,\um), the original uncorrected spectrum is retained
with the telluric transmission column set to NaN.

Wavelengths in the output spectra are updated using the \viper{} fit
where available. For detector orders without a successful fit (due to weak
telluric features, the CO$_2$-saturated region, or low signal-to-noise),
the wavelength scale is corrected by fitting a linear velocity offset
$\Delta v(\lambda)$ to the differences between the \viper{} and
reference wavelengths of all well-fitted orders (percentage residual RMS
$< 30$) across all three detector chips simultaneously. The three chips
are at fixed relative positions in the focal plane, so the wavelength
correction varies smoothly across the full spectral range and pooling all
chips provides a well-constrained fit even when individual chips have few
fitted orders. The resulting correction, typically 1--5\,\kms{}, is
applied to the reference wavelength polynomial from the tracing table.
A correction exceeding 100\,\kms{} is considered unphysical and the
order retains the reference wavelength.

In an earlier attempt, we used a 2D polynomial surface $\lambda(x,\,
\mathrm{order})$, fit independently per chip, to interpolate absolute
wavelengths to unfitted orders. This approach proved unreliable:
with only two or three fitted orders per chip the surface was
underconstrained, and quadratic extrapolation to edge orders produced
wavelengths off by factors of several. Fitting a velocity \emph{correction}
across all chips avoids this because the corrections are small and smooth,
keeping extrapolation errors limited.

The precision of the \viper{} wavelength solution on science data is
assessed using four complementary metrics (Fig.~\ref{fig:wlprecision}).
Since \viper{} fits each detector segment independently, the rigid
mechanical coupling between the three \crires{} detectors provides a
built-in consistency check: the wavelength gap between chip centres at a
given echelle order is a physical constant, so its scatter across
observations measures the fit repeatability. Analogous gaps between
adjacent orders on the same chip provide an independent check. Most
directly, the A and B nod positions of each observation receive
independent \viper{} fits to the same telluric spectrum; their
wavelength difference at a given pixel should be constant (reflecting the
slit tilt) with scatter determined only by fit noise.

We evaluate the wavelength at detector centre (pixel~1024) for each
fitted segment in all 4068 science observations with at least one order
passing the quality filter ($\mathrm{prms} < 10$), and compute the
deviation of each measurement from the median across all observations of
the same wavelength setting, chip, and order. The A--B comparison yields
the tightest constraint, with a median per-group RMS of 253\,m\,s$^{-1}$
across 218 well-behaved groups. Since each A--B difference combines the
noise of two independent fits, the implied single-fit precision is
$253 / \sqrt{2} \approx 179$\,m\,s$^{-1}$.  The inter-chip and
inter-order gap metrics give median RMS values of 292 and
659\,m\,s$^{-1}$, respectively, while the absolute wavelength
repeatability (which additionally includes real observation-to-observation
variations from atmospheric and instrumental changes) has a median RMS of
${\sim}2.2$\,\kms. These results apply to orders with sufficient telluric
line density for a well-constrained fit; orders near atmospheric
transmission windows or the CO$_2$ fundamental band show significantly
larger scatter.

\begin{figure}
  \centering
  \includegraphics[width=\columnwidth]{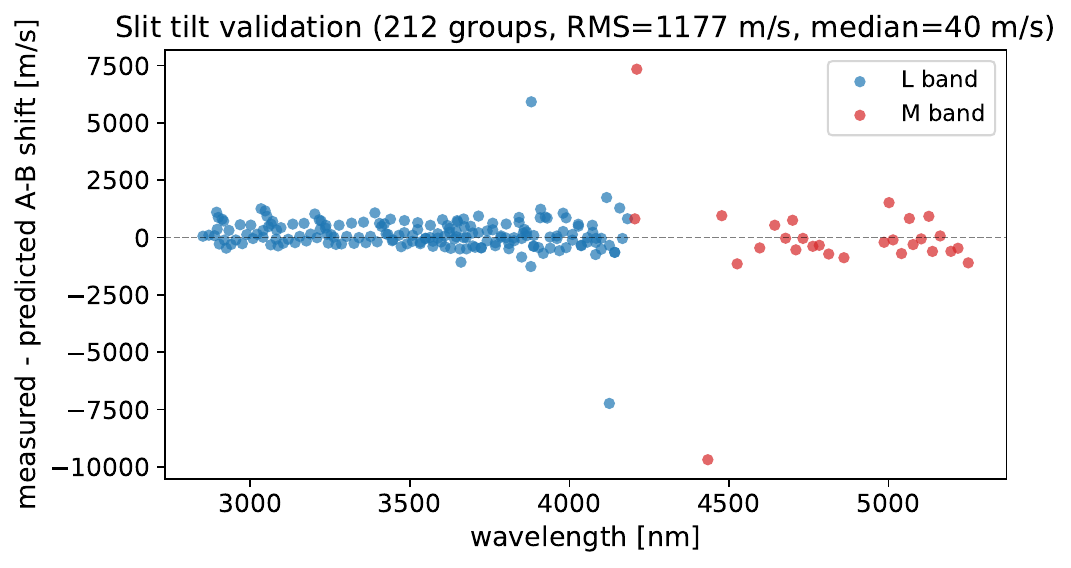}
  \caption{Residual between the measured median A--B wavelength shift and
    the shift predicted by the slit tilt model, as a function of
    wavelength.  Blue and red points denote L- and M-band groups,
    respectively.  The RMS residual of 1177\,m\,s$^{-1}$ and median
    offset of 40\,m\,s$^{-1}$ are consistent with a correctly measured
    tilt.}
  \label{fig:tiltcomp}
\end{figure}

As an independent validation of the slit tilt calibration, we compare the
measured A--B wavelength shifts to the predictions from the
\texttt{SlitPolyB} tilt coefficients stored in the tracing tables.  For
each observation the predicted velocity shift is $\Delta v = t \cdot
\Delta{}y \cdot (\mathrm{d}\lambda/\mathrm{d}x) / \lambda \cdot c$,
where $t$ is the tilt in pixels per spatial pixel (from the linear fit described in
Sect.~\ref{sec:tilt}), $\Delta{}y$ the nod throw in pixels (from
the \texttt{ESO SEQ NODTHROW} header converted
via the 0\farcs059\,pixel$^{-1}$ plate scale),
$\mathrm{d}\lambda/\mathrm{d}x$ the local dispersion, and $\lambda$ the
wavelength, both from the tracing table polynomial.  Across 212 (setting, chip, order) groups
the residuals between the measured and predicted A--B shifts have an RMS
of 1177\,m\,s$^{-1}$ and a median offset of only 40\,m\,s$^{-1}$
(Fig.~\ref{fig:tiltcomp}), perfectly in line with a correctly measured
tilt.

\begin{figure}
  \centering
  \includegraphics[width=\columnwidth]{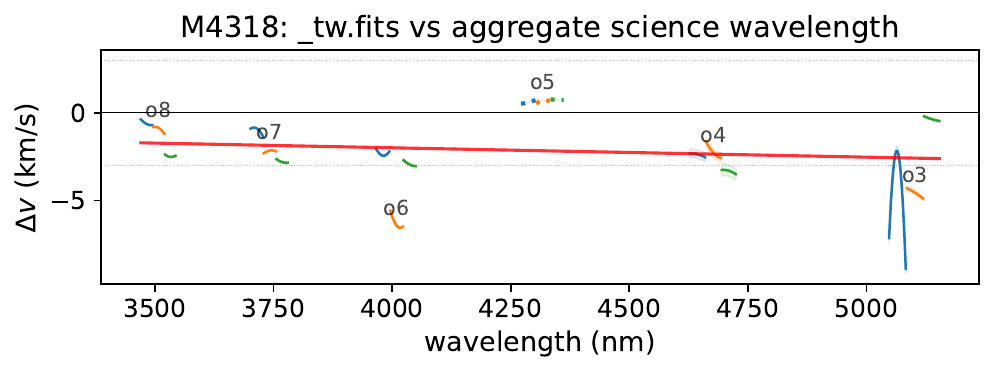}
  \caption{Velocity offset between the reference tracing table
    wavelengths and the median science \viper{} solution for the M4318
    setting, shown as a representative example. Each curve segment
    corresponds to one detector order, coloured by chip (blue: CHIP1,
    orange: CHIP2, green: CHIP3). Dotted segments mark orders that retained
    pipeline wavelengths; solid segments have \viper{} wavelengths from
    the telluric standard. The red line is the linear fit used to correct
    the tracing table. The shaded bands show the uncertainty of the
    median.}
  \label{fig:twcorr}
\end{figure}

Once all science spectra are processed, we use the aggregate
\viper{} wavelength solutions to refine the reference tracing tables
themselves. For each wavelength setting, we compare the default
wavelength polynomial to the median science \viper{} solution per
detector order, evaluated at pixel~1024 from all observations passing
the quality filter. The resulting velocity offsets follow a smooth linear
trend with wavelength within each setting, reaching $-$2 to
$-$8\,\kms{} in L~band and up to $+$8\,\kms{} in some M-band settings
(Fig.~\ref{fig:twcorr}).
We fit a linear relation $\Delta v = a + b\,\lambda$ per setting
(iterative $3\sigma$ clipping, typical RMS of the fit $\lesssim
1$\,\kms) and apply it as a multiplicative correction to all three
wavelength polynomial coefficients. A linear velocity offset produces
corrections not only to the zero-point~$c_0$ but also to the
dispersion~$c_1$ and curvature~$c_2$, because $\lambda_\mathrm{new}(x)
= \lambda(x)\,[1 - \Delta v(\lambda) / c]$ and $\Delta v$ itself
depends on~$\lambda(x)$.  The corrected tracing tables are provided as
the reference calibration products; they are not used to re-reduce the
archive spectra presented here, but serve as improved starting points for
future reductions of data in these wavelength settings.

\subsection{Caveats}
\label{sec:caveats}

We believe this data release is a significant improvement over what
was hitherto available, for the vast majority of the included observations.
One size, however, does not fit all, and there are a few caveats to be aware of.

The reductions carried out here assume there is a single target at two 
slit positions A and B separated by the nod throw. 
The spectra get extracted there, collapsing
all light in the extraction window of fixed height (45 pix) into a single spectrum.
Naturally, this
is not useful for observations where spatial resolution along the slit is relevant,
for example extended targets or binary stars with both companions in
the slit.

We emphasise that \viper{} is used here primarily as a wavelength-calibration
tool: the forward-model fit pins the wavelength scale to the forest of
atmospheric lines in each order, which is how we recover the L/M wavelength
solution in the absence of usual wavelength calibration frames: Uranium-Neon and Fabry-Perot. 
The byproduct telluric model is usable for initial inspection and works acceptably on our data 
(see e.g.\ Sect.~\ref{sec:examples}), but it is not
intended as a science-grade correction. Users with demanding
telluric-cancellation needs should run their own tailored fits
(e.g.\ \texttt{molecfit}) against the uncorrected spectrum, which is always
recoverable as $\texttt{SPEC} \times \texttt{TELLUR}$.

Additionally, spectral features that are both strong and so wide that they dominate
the spectrum in the detector order that contains them, for example Br$\alpha$, can throw off
the continuum fit of \viper\, in the sense that the continuum gets wrongly
pulled to match part of the line. This is generally quite obvious from the telluric model and the
residuals, and needs to be remedied by re-fitting the continuum without
the spectral feature.

As mentioned above (\S \ref{sec:extraction}),
 data taken without metrology can be partially cut off by the flat-field.
 
No non-linearity correction was applied to the spectra, cf.~\S \ref{sec:calibs}.

\subsection{Use of AI tools}
\label{sec:ai}

Claude Opus 4.6 \citep{claude_opus46} was used extensively in this project --
it would not have happened without it, since the AI collaboration turned weeks
of work into days.

With close supervision, Claude Code wrote and executed the scripts that carry out the aforementioned steps:
\begin{itemize}
\item query the ESO archive for all LM data and build a local SQLite database with meta-information
\item retrieve the data from the ESO archive
\item group data by nodding AB pairs and associate the nearest flat-field
\item generate the SOF with filenames and tags for the \crtwores\ recipes
\item measure the slit tilt from the telluric stars' A-B difference
\item find the best shift in Y for the trace polynomials
\item orchestrate the parallel runs of 10,000 pipeline reductions
\item install and run \texttt{molecfit} to generate the input atmospheric model for \viper
\item carry out the telluric fits with \viper\ and plot the results
\item aggregate the telluric fit results from all data to verify the slit tilt
\item analyse the wavelength precision by comparing the telluric fits from all observations
\end{itemize}

In addition, the web-application that runs the data archive is entirely 
AI-generated.

The authors are confident in the methodological correctness of the applied
concepts and when inspecting the code find no reason to doubt
the results.

The first draft of some sections of this paper was also written in collaboration with Claude, often
as the summary of the analysis session that produced the results.
The text was then edited and checked for correctness by the authors.


\begin{figure}
  \centering
  \includegraphics[width=\columnwidth]{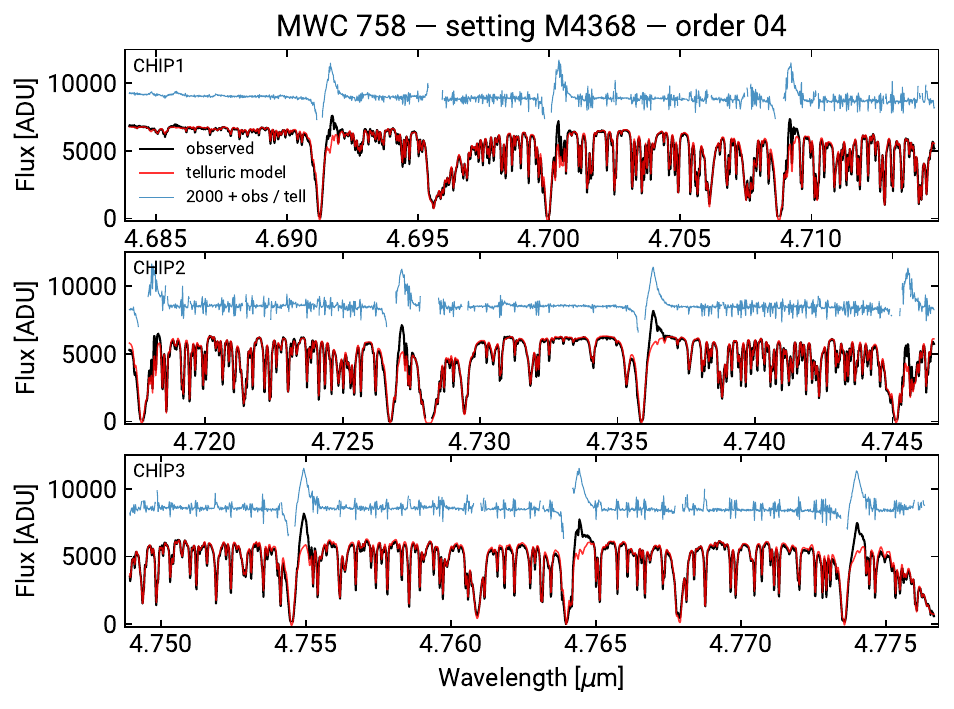}
  \caption{$^{12}$CO $v=1\!-\!0$ fundamental P-branch toward the Herbig\,Ae
    transition-disk star MWC\,758 (M4368, order 04, spanning the three
    detector chips). Each panel shows the observed spectrum (black) with the
    \viper\ forward model of the telluric transmission times the
    fitted stellar continuum overlaid (red), plus the telluric-corrected
    spectrum (blue, offset vertically for clarity). Narrow CO emission peaks
    from the inner disk stand out cleanly above the atmospheric absorption
    after correction.}
  \label{fig:mwc758_CO}
\end{figure}

\begin{figure}
  \centering
  \includegraphics[width=\columnwidth]{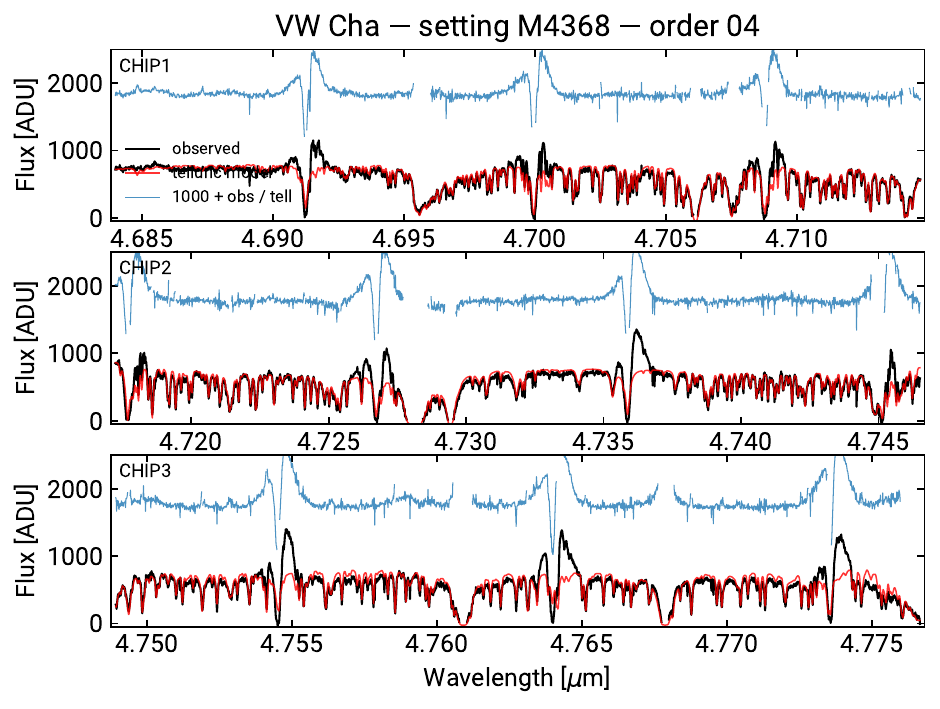}
  \caption{Same as Fig.~\ref{fig:mwc758_CO} but for the classical T Tauri
    star VW\,Cha.}
  \label{fig:vwcha_CO}
\end{figure}

\begin{figure}
  \centering
  \includegraphics[width=\columnwidth]{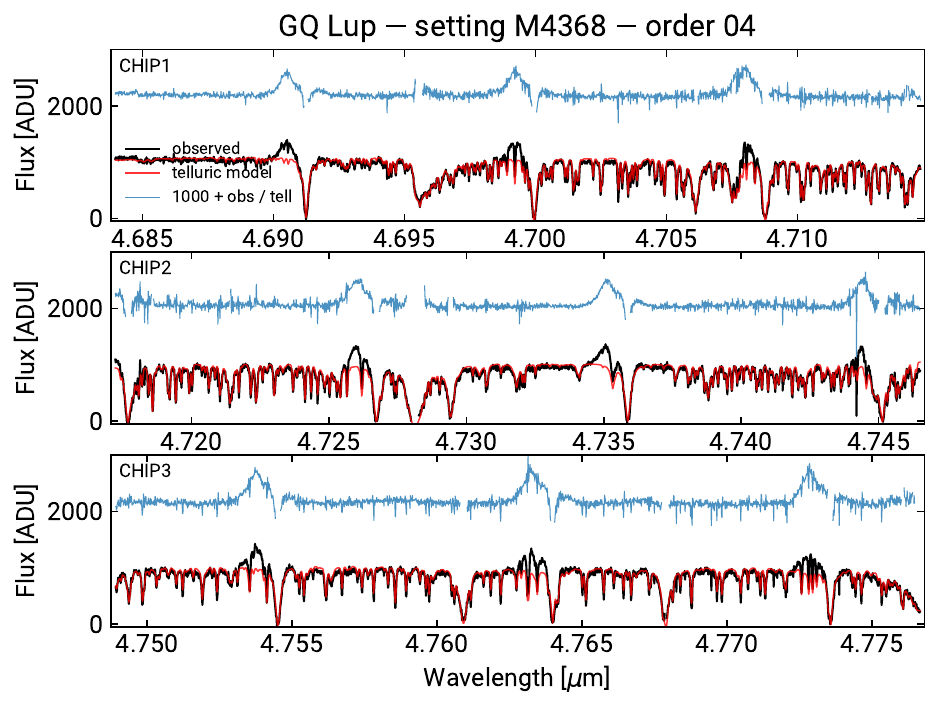}
  \caption{Same as Fig.~\ref{fig:mwc758_CO} but for the classical T Tauri
    star GQ\,Lup.}
  \label{fig:gqlup_CO}
\end{figure}

\section{The data archive website and code repository}
\label{sec:archive}

The reduced spectra are served through a web application at
\url{https://www.astro.uu.se/crires-lm/}. The front page presents a
filterable and sortable table of all observations for which telluric-corrected
spectra are available, covering all 5649 nod pairs. Users can filter by target
name, wavelength setting, or ESO programme ID.

Each observation page shows an interactive spectrum plot (using Plotly)
displaying the spectrum and the telluric model. Alongside, there are
several static plots that allow a quick first assessment of the data:
\begin{itemize}
\item The $|A-B|$ image for the 3 detectors, with the extraction window indicated as black/white dashed lines.
\item Spectrum and telluric model for each order, plus residual.
\item Wavelength correction, i.e.~the velocity offset of each detector-order from the default wavelength solution. 
 The black line, if present, indicates the fit to the good solutions to extrapolate to the indicated orders that did not get good telluric fits on the data.
\end{itemize}

FITS files are available for download, containing the corrected spectra;
 the output columns per spectral order include the corrected
spectrum, error, wavelength, telluric transmission, and fitted continuum.
All information needed to reconstruct the
original spectrum is present: the \texttt{SPEC} column contains the telluric-corrected
spectrum (the observed spectrum divided by the telluric model), while the
fitted telluric transmission and continuum are stored in separate
\texttt{TELLUR} and \texttt{CONT} columns. The original uncorrected spectrum
can thus be recovered as $\texttt{SPEC} \times \texttt{TELLUR}$, and the full
\viper\ forward model as $\texttt{CONT} \times \texttt{TELLUR}$.

Observation pages also link to the page of the flat-field that was used for the
reduction, which shows the normalized flat-field, the blaze functions and meta-data.

Lastly, a table lists all raw frames that went into the data presented at the current page.
Clicking on an individual AB pair leads to the reduction page for that specific AB pair alone.

The full reduction code, scripts, static input data, recipe parameter files, and the webapp are
 available from the repository at 
\url{https://github.com/ivh/CRIRES-LM}. Reproducing this work should be as
straightforward as re-running the download and reduction scripts in the documented
order. On a machine with 32 cores, using GNU \texttt{parallel} \citep{tange_parallel}, this takes
on the order of a day.

Also in the repository, there are the updated trace table files
(\texttt{*\_tw.fits}) with cleaned-up traces, slit-tilt values and improved default
wavelengths. These can be used as drop-in
replacements for other data reductions, and are offered to ESO in the
hope that they will ship with the next release of the cr2res pipeline. 

Part of the motivation for making this an arXiv-only paper is to allow for
easy future updates, both in terms of adding data as more becomes
public, and in terms of code and methods for when further improvements
are found. The repository has a \emph{git tag} matching the arXiv-version of
this paper, and we will add a change-log here for versions after this initial one.

To cite this work and the data, please use the arXiv number
 \texttt{astro-ph/YYMM.NNNNN}, along with \citet{crires_lm_data} for the
 reprocessed spectra (a ZIP archive on Zenodo) and \citet{crires_lm_code}
 for the code release matching this version of the paper.

\section{Showcase examples}
\label{sec:examples}

The purpose of this work is not to analyse the improved spectra for scientific results,
but to enable other researchers with the necessary expertise to more easily exploit
these hitherto neglected observations. Nevertheless, here are a few examples of
what caught our eye when browsing the data.

\begin{figure}
  \centering
  \includegraphics[width=\columnwidth]{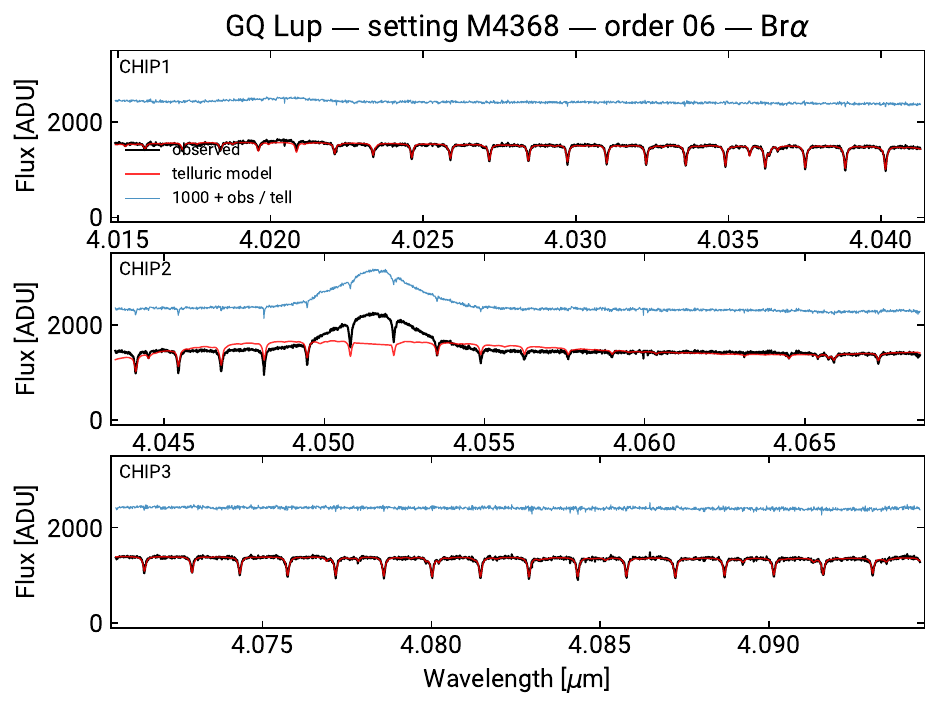}
  \caption{Br$\alpha$ at 4.052\,\um{} in GQ\,Lup, in
    order 06 of the same M4368 observation as Fig.~\ref{fig:gqlup_CO}.}
  \label{fig:gqlup_winds}
\end{figure}

\begin{figure}
  \centering
  \includegraphics[width=\columnwidth]{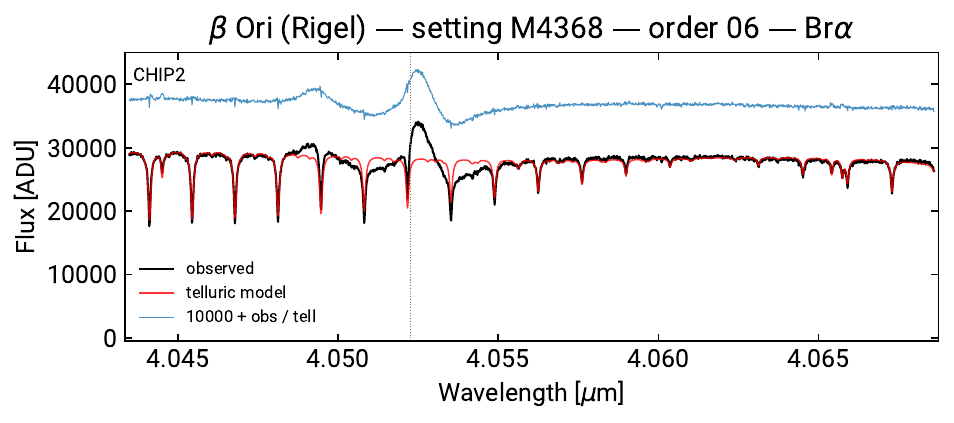}
  \caption{Br$\alpha$ at 4.052\,\um{} in strong emission
    in the spectrum of $\beta$\,Ori (Rigel)}
  \label{fig:betOri_Bra}
\end{figure}

A striking feature in the M band is the $^{12}$CO $v=1\!-\!0$
fundamental ro-vibrational band near 4.7\,\um{}, which falls into the M4368
setting. In the star MWC~758, the CO is seen in emission on top of the deep
telluric forest\footnote{\url{https://www.astro.uu.se/crires-lm/obs/MWC_758_M4368_2024-12-31_02310}} 
(Fig.~\ref{fig:mwc758_CO}). Similar CO emission is visible in
the young T Tauri stars 
VW~Cha\footnote{\url{https://www.astro.uu.se/crires-lm/obs/V_VW_Cha_M4368_2023-04-28_00032}} 
(Fig.~\ref{fig:vwcha_CO}) 
and GQ~Lup\footnote{\url{https://www.astro.uu.se/crires-lm/obs/V_GQ_Lup_M4368_2024-03-16_09003}}
(Fig.~\ref{fig:gqlup_CO}). GQ~Lup additionally shows Br$\alpha$ at
4.052~\um{} as a tracer of magnetospheric accretion \citep[e.g.][]{komarova2020}
as shown in Fig.~\ref{fig:gqlup_winds}.
The same Br$\alpha$ line is detected in strong emission in the supergiant
$\beta$\,Ori\footnote{\url{https://www.astro.uu.se/crires-lm/obs/bet_Ori_M4368_2024-12-09_02033}} 
(Rigel, Fig.~\ref{fig:betOri_Bra}).

\begin{figure}
  \centering
  \includegraphics[width=\columnwidth]{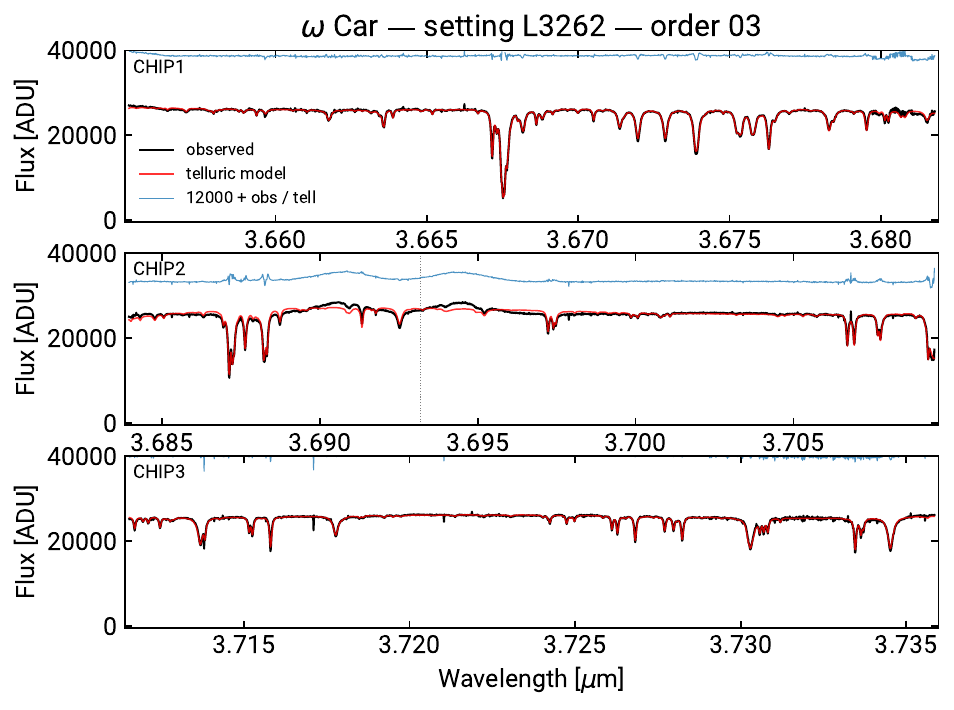}
  \caption{A Humphreys recombination line at
    3.693\,\um{} in emission in the spectrum of
    $\omega$\,Car.}
  \label{fig:omeCar_Hu18}
\end{figure}

The L-band also offers access to the Humphreys series, a forest of transitions between
3.3 and 4.0\,\um{} \citep[e.g.][]{lenorzer2002}. Fig.~\ref{fig:omeCar_Hu18} shows one such line 
in emission toward the star $\omega$\,Car\footnote{\url{https://www.astro.uu.se/crires-lm/obs/ome_Car_L3262_2023-12-24_07335}}. 

All of these spectra can be browsed and downloaded from the project website.

\section{Summary}
\label{sec:summary}

We reprocessed all 11\,131 public \crires{} L- and M-band science
frames from the ESO archive into 5649 telluric-corrected nod-pair spectra
covering 156 targets across 16 wavelength settings. The reprocessing
includes slit tilt from telluric standard fits, and wavelength calibration
from per-observation telluric correction.
The resulting archive is browsable and downloadable at
\url{https://www.astro.uu.se/crires-lm/}.

\begin{acknowledgements}
We thank N.~Piskunov for early feedback on this project and the draft text; and for
his long-standing role in the \crires\ project and its data reduction pipeline.

Based on observations made with ESO Telescopes at the La Silla Paranal Observatory under programme ID(s)
60.A-9801(B),
107.22SP.001,
107.22T7.001,
107.22T7.002,
107.22TZ.001,
107.22UB.001,
108.223N.001,
108.225Y.003,
108.228B.001,
108.228B.002,
108.22JJ.001,
109.231P.001,
109.2320.001,
109.23BJ.001,
109.23FT.001,
109.23GX.002,
109.23GX.004,
109.23GX.006,
110.23NC.001,
110.244F.001,
110.244F.002,
110.249Y.004,
111.24V2.001,
111.24VY.001,
111.24VY.002,
111.24XW.003,
111.250R.001,
111.250R.002,
111.250Y.001,
111.250Y.002,
111.254J.004,
112.25MR.001,
112.25MU.001,
112.25MU.002,
112.25PX.001,
112.25PX.002,
112.25PX.003,
112.26WN.001,
112.26WN.002,
113.269C.002,
113.269T.001,
113.26AX.002,
113.26F9.001,
113.26PE.001,
113.26VH.001,
114.271M.001,
114.271M.002,
114.271M.003,
114.271M.004,
114.279K.002,
114.279T.001,
114.279X.001,
114.27LL.002,
114.27PB.001,
114.27PB.002.
\end{acknowledgements}

\bibliographystyle{aa}
\bibliography{crires_lm}

\end{document}